\documentclass[12pt,epsfig,amsfonts]{article}
\usepackage{graphicx}
\renewcommand{\bottomfraction}{0.7}

\newcommand{\ci}[1]{\cite{#1}}
\newcommand{\bi}[1]{\bibitem{#1}}

\newcommand{\ba}{\begin{eqnarray}}
\newcommand{\ea}{\end{eqnarray}}
\newcommand{\beqs}{\begin{eqnarray}}
\newcommand{\eeqs}{\end{eqnarray}}

\begin{document}
\title{GPDs of the nucleons \\ and elastic scattering at high energies}

\author{  O.V. Selyugin\footnote{selugin@theor.jinr.ru}\\[1ex]
 \it BLTP, Joint Institute for Nuclear Research,
 }



\date{}
\maketitle

\begin{abstract}
  Taking into account the electromagnetic and gravitational form factors, calculated from a new set of $t $-dependent GPDs, a new model is built. The real part of the hadronic amplitude is determined only through complex $s$. In the framework of this model the quantitative  description of all existing experimental data at $52.8  \leq \sqrt{s} \leq  1960 \ $GeV, including the Coulomb range and large momentum transfers ($0.0008 \leq |t| \leq  9.75 \ $GeV$^2$), is obtained with only $3$  fitting  high energy parameters.
  The comparison with the preliminary data of the TOTEM Collaboration at an energy of $7$ TeV is made.
\end{abstract}

\section{Introduction}
	The dynamics of strong interactions  finds its most
  complete representation in elastic scattering at small angles.
  Only in this region of interactions we can measure the basic properties of
  the non-perturbative strong interaction which
  define the hadron structure: the total cross section,
  the slope of the diffraction peak and the parameter $\rho(s,t) $.
  Their values
  are connected, on the one hand, with the large-scale structure of hadrons and,
  on the other hand, with the first principles which lead to the
  theorems on the behavior of the scattering amplitudes at asymptotic
  energies \ci{mart,roy}.

There are indeed many different models for the description of hadron elastic
 scattering at small angles \cite{Rev-LHC,TOTEM-1395}.  They lead to  the different
 predictions for the structure of the scattering amplitude at asymptotic
 energies, where the diffraction  processes can display complicated
 features \cite{dif04}.  This concerns especially the asymptotic unitarity
 bound connected with the Black Disk Limit (BDL) \cite{CPS-EPJ08}.
   In Chow-Yang model \cite{WY-65,CY-68} it was assumed that the hadron interaction  to be    proportional the overlapping of the matter distribution of the hadrons and in Wu and Yang  \cite{WY-65} suggested that the matter distribution  is proportional to the charge distribution of the hadron.
 Then many model were used  the electromagnetic form factors of hadron,
but, in most part they change his form to describe the experimental data,
 as was made in  the famous  Bourrely-Soffer-Wu model \cite{BSW}.
 The parameters of the obtained form-factor are determined by the fit of the differential cross sections.

 Now we present a model that used two form factors
  determined by one   function - generalized parton distributions (GPDs).
  Following  sum rules  valid for the momentum of the GPDs \cite{Ji97,R97} the integration over the momentum fraction, $x$, yields the conventional
  electromagnetic form factors and the integration of the next momentum of the GPDs \cite{Ji97,ST-FF} yields
   the gravitation form factors.
    The correlation between hadron form factors and energy momentum tensor were discussed long time ago \cite{Pagels}
    and recently \cite{Polyakov,AC08}.
    So,  both the form factors are independent of the fitting procedure of the differential cross sections. Note that the form of the GPDs is determined, on  the one hand, by   the deep-inelastic processes and, on the other hand, by the measure of the electromagnetic form factor from  the electron-nucleon elastic scattering. Hence, the form of the electromagnetic form factor (first momentum of GPDs) determines the form of the second form factor (second momentum of GPDs). This scheme is supported by the good description of the experimental data in the Coulomb-hadron interference region and large momentum transfer at high energies of one amplitude with a few free parameters.

   The impact of the hard pomeron contribution on the elastic differential cross sections is very important for  understanding the properties of the QCD in the non-perturbative regime \cite{mrt}. Note, that the real part of the hard pomeron is essentially large then   the real part of the soft pomeron.  Now in \cite{DL-11hp} it is suggested that such a contribution can be explained by the preliminary result
   of the TOTEM Collaboration \cite{TOTEM-111008a} on the elastic proton-proton differential cross sections.   In our model, the real part of the hadronic amplitude is determined only through complex $s$ satisfying the cross symmetric relation.  In the framework of this model, the quantitative  description of all  existing experimental data at $52.8 \leq \sqrt{s} \leq 1960  $ GeV, including   the Coulomb range and large momentum transfers $0.0008 \leq |t| \leq  9.75 \ $GeV$^2$  , is obtained with only   $3$ fitting high energy parameters. The comparison of the predictions of the model  at  $7$ TeV and preliminary data of the TOTEM collaboration are shown to coincide well.
There is some small place, especially in the region of the diffraction dip, for the small correction contributions which are  determined by the odderon,    and possibly the   spin-dependent part of the scattering amplitude which gives a small contribution at large momentum transfer.
 In the framework of the model, only the Born term of the scattering amplitude
   is introduced. Then the whole scattering amplitude is obtained as a result of the unitarization procedure  of the hadron Born term that is then  summed with the Coulomb term. The Coulomb-hadron interference phase is also taken into account.    The essential moment of the model is that both parts of the Born term
  of the scattering amplitude have the positive sign,  and the diffraction structure is determined by the unitarization procedure.

   The  electromagnetic and hadronic parts of the elastic
   scattering amplitudes used in the model are presented in the second and third sections.
   In the fourth section, we introduce the  hadron form factors obtained from the first and second momenta of GPds.
   Our fitting procedure and the description of the high energy differential cross sections of the
   proton-proton and proton-antiproton scattering     are presented in the fifth and sixth sections.
   Also, we stretch our model on $7$ TeV and compare the model calculations with the preliminary data
   of the TOTEM Collaboration.  Then, the model calculations are compared
   with the experimental data at low energies $\sqrt{s} = 30 \ $GeV.
   In the seventh section and in the conclusion, we show and discuss the model calculations for the
   total cross sections and the value of $\rho(s,t)$ obtained in the framework of the model.

\section{Electromagnetic part of the hadron scattering amplitude}

  The differential cross
  sections of nucleon-nucleon elastic scattering  can be written as the sum of different
  helicity  amplitudes:
\begin{eqnarray}
  \frac{d\sigma}{dt} =
 \frac{2 \pi}{s^{2}} (|\Phi_{1}|^{2} +|\Phi_{2}|^{2} +|\Phi_{3}|^{2}
  +|\Phi_{4}|^{2}
  +4 | \Phi_{5}|^{2} ). \label{dsdt}
\end{eqnarray}
\renewcommand{\bottomfraction}{0.7}
  The total helicity amplitudes can be written as $\Phi_{i}(s,t) =
  F^{h}_{i}(s,t)+F^{\rm em}_{i}(s,t) e^{\varphi(s,t)} $\,, where
 $F^{h}_{i}(s,t) $ comes from the strong interactions,
 $F^{\rm em}_{i}(s,t) $ from the electromagnetic interactions and
 $\varphi(s,t) $
 is the interference phase factor between the electromagnetic and strong
 interactions \cite{selmp1,selmp2,Selphase}.
 For the hadron part the amplitude with spin-flip is neglected in this approximation, as usual at high energy.


 The electromagnetic amplitude can be calculated in the framework of QED.
    In the high energy approximation, it can be  obtained \cite{bgl}
  for the spin-non-flip amplitudes:
  \begin{eqnarray}
  F^{em}_{1}(t) = \alpha f_{1}^{2}(t) \frac{s-2 m^2}{t}; \ \ \  F^{em}_3(t) = F^{em}_1;
  \end{eqnarray}
   and for spin-flip amplitudes:
 \begin{eqnarray}
  F^{em}_2(t) =  \alpha  \frac{f_{2}^{2}(t)}{4 m^2} s; \ \ \
   F^{em}_{4}(t) =  - F^{em}_{2}(t), \ \  \  \\ \nonumber
  F^{em}_5(t) =  \alpha \frac{s }{2m \sqrt{|t|}} f_{1}(t) \ f_{2}(t),
  \end{eqnarray}
  where the form factors are:
    \begin{eqnarray}
    f_{1}(t) = \frac{4 m_{p}^{2} - (1+k) \ t}{ 4 m_{p}^{2} - \ t} \ Gd(t); \\ \nonumber
    f_{2}(t) = \frac{4 m_{p}^{2} \ k}{ 4 m_{p}^{2} - \ t} \ Gd(t);
\end{eqnarray}
  with $k$ relative to the anomalous magnetic moment, and $G_d(t)$ has the conventual dipole form
 \begin{eqnarray}
   Gd(t)= 1/(1-t/0.71)^2.
  \end{eqnarray}
  \section{Main hadronic amplitude}
      The model is based on the representation that at high energies a hadron interaction in the non-perturbative regime
      is determined by the reggeized-gluon exchange. The cross-even part of this amplitude can have two non-perturbative parts, possible standard pomeron - $P_{2np}$) and cross-even part of the three non-perturbative gluons ($P_{3np}$) case.
      The interaction of these two objects is proportional to two different form factors of the hadron.
      This is the main assumption of the model. Of course, we cannot insist on the origin of the second
      term of the scattering amplitude. It can be of a different nature. However, in any case, it has the cross-even properties and positive sign. The second important assumption is that we chose the slope of the second term
       $4$ time smaller than a slope of the first term, by  analogy with the two pomeron cut.
      Both terms have the same intercept.

      The form factors are determined by the General parton distributions of the hadron (GPDs).
      The first form factor, corresponding to the first momentum of GPDs is the conventional electromagnetic
      form factor - $G(t)$. The second form factor is determined by the second momentum of GPDs -$A(t)$.
      The parameters and $t$-dependence of the GPDs are determined by the standard parton distribution
      functions, so by the experimental data on the deep-inelastic scattering, and by the experimental data
      for the electromagnetic form factors (see \cite{ST-PRDGPD}).

      Hence, the Born term of the elastic hadron amplitude can be written as
  \begin{eqnarray}
 F_{h}^{Born}(s,t) \ = && h_1 \ G^{2}(t) \ F_{a}(s,t) \ (1+r_1/\hat{s}^{0.5}) \\ \nonumber
                  + && h_{2} \  A^{2}(t) \ F_{b}(s,t) \ (1+r_2/\hat{s}^{0.5})
\end{eqnarray}
  where $F_{a}(s,t)$ and $F_{b}(s,t)$  has the standard Regge form 
  \begin{eqnarray}
 F_{a}(s,t) \ = \hat{s}^{\epsilon_1} \ e^{B(s) \ t}; \ \ \
 F_{b}(s,t) \ = \hat{s}^{\epsilon_1} \ e^{B(s)/4 \ t},
\end{eqnarray}
 with $G(t)=G_E(t)$ is the Sachs electric form factor relative to the first moment of GPDs and $A(t)$ relative to the second moment of GPDs.
\begin{eqnarray}
G(t) &=&\frac{L_{1}^{4}}{(L_{1}^2-t)^2} \ \frac{4m_{p}^{2}- (1+k) \ t}{4m_{p}^{2}- \ t} \\
A(t)&=& \frac{L_{2}^4}{(L_{2}^2-t)^2}
 .\label{overlap}
 \end{eqnarray}
with the parameters:  $L^{2}_{1}=0.71 \ $GeV$^2$; $L_{2}^2=2 $~GeV$^2 $.
 \begin{eqnarray}
    \hat{s}=s \ e^{-i \pi/2}/s_{0} ;  \ \ \ s_{0}=1 \ {\rm GeV^2}.
\end{eqnarray}
  The slope of the scattering amplitude has the standard logarithmic dependence on the energy.
 \begin{eqnarray}
    B(s) = \alpha^{\prime} \ ln(\hat{s}) .
\end{eqnarray}
  with $\alpha^{\prime}=0.24$GeV$^{-2}$.

  \begin{table*} 
\label{tab:2}       
\begin{center}
\begin{tabular}{c|c|c||c|c}
\hline\noalign{\smallskip}
 \multicolumn{3}{c} { high energy parameters }& \multicolumn{2}{c}{low energy parameters}  \\
\noalign{\smallskip}\hline\noalign{\smallskip}
 $h_1$, GeV$^{-2}$  & $h_2$, GeV$^{-2}$ & $\epsilon_1 $  & $r_1$, GeV & $r_2$, GeV   \\
 $1.03 \pm 0.02 $  & $3.31 \pm 0.02 $ & $0.11 \pm 0.01 $ & $ 11.95 \pm 0.5 $ &$ -5.9 \pm 0.8$  \\
\noalign{\smallskip}\hline
\end{tabular}
\caption{The basic parameters of the model are determined by fitting experimental data.}
\end{center}
\end{table*}

\begin{figure}
\begin{center}
 \includegraphics[width=0.4\textwidth]{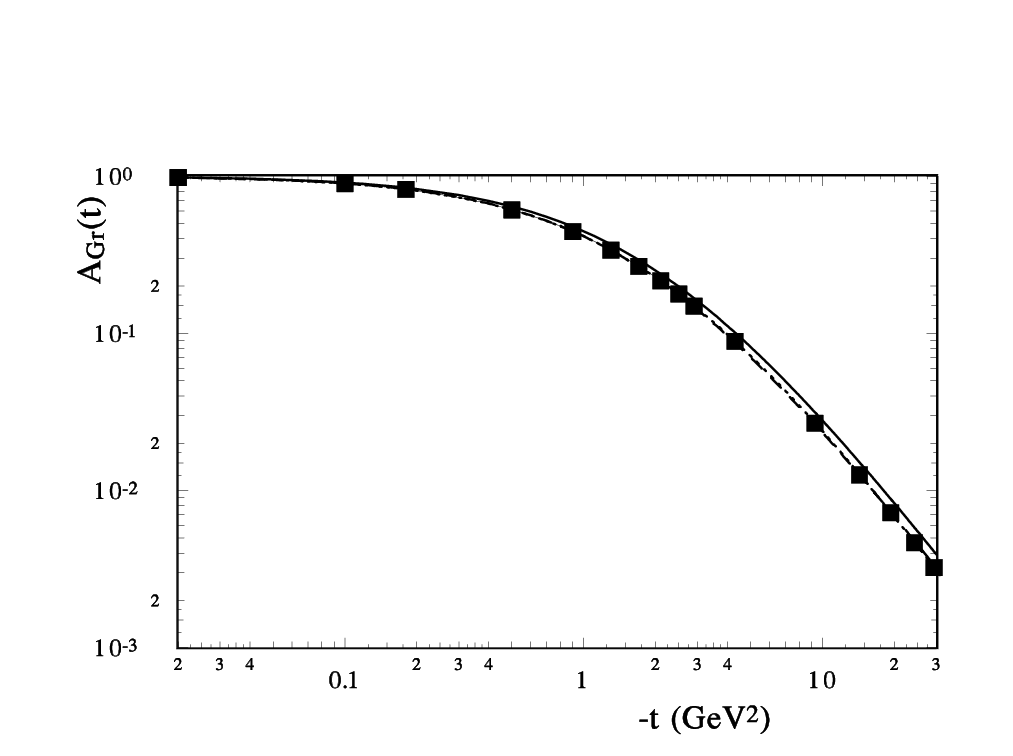}
\end{center}
\caption{
 The hadron form factor $A(t)$ - (the second momentum of GPDs) in the $t$ representation (
   squares - the numerical calculation of the integral over $x$ is normalized to $1$, hard line - the dipole form with $L^2 = 2.0 \ $GeV$^2$).
   dashed line - the dipole form  $L^2 = 1.8 \ $GeV$^2$).
      }
\label{Fig_2}
\end{figure}

\label{sec:figures}
\begin{figure*}
\begin{center}
\includegraphics[width=0.4\textwidth] {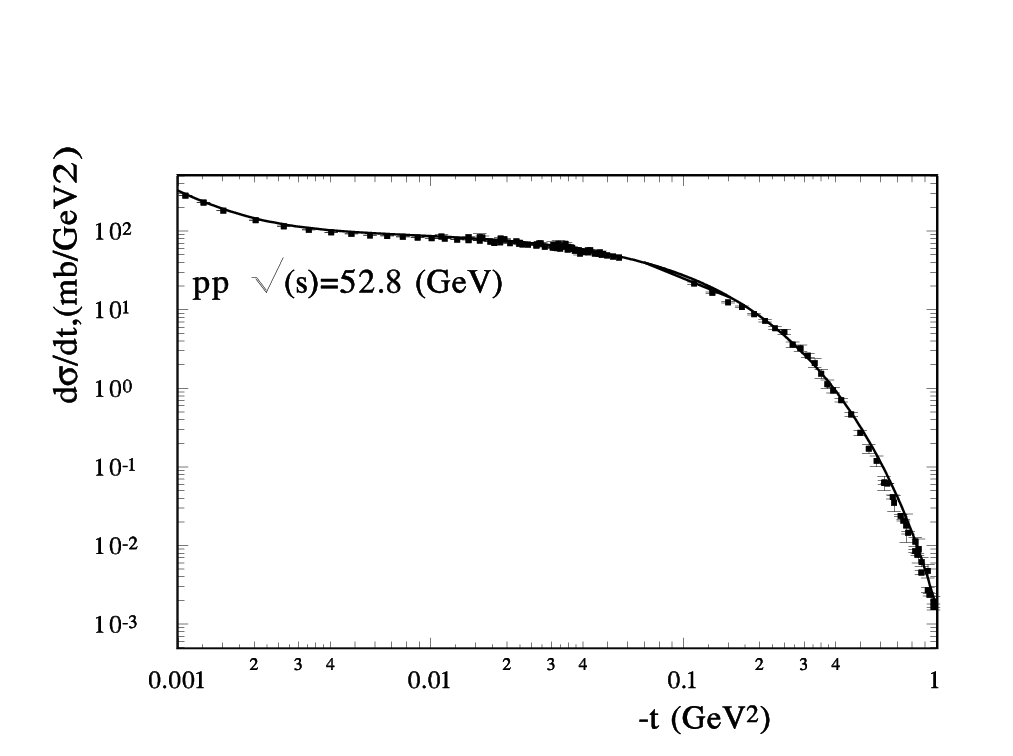}
\includegraphics[width=0.4\textwidth] {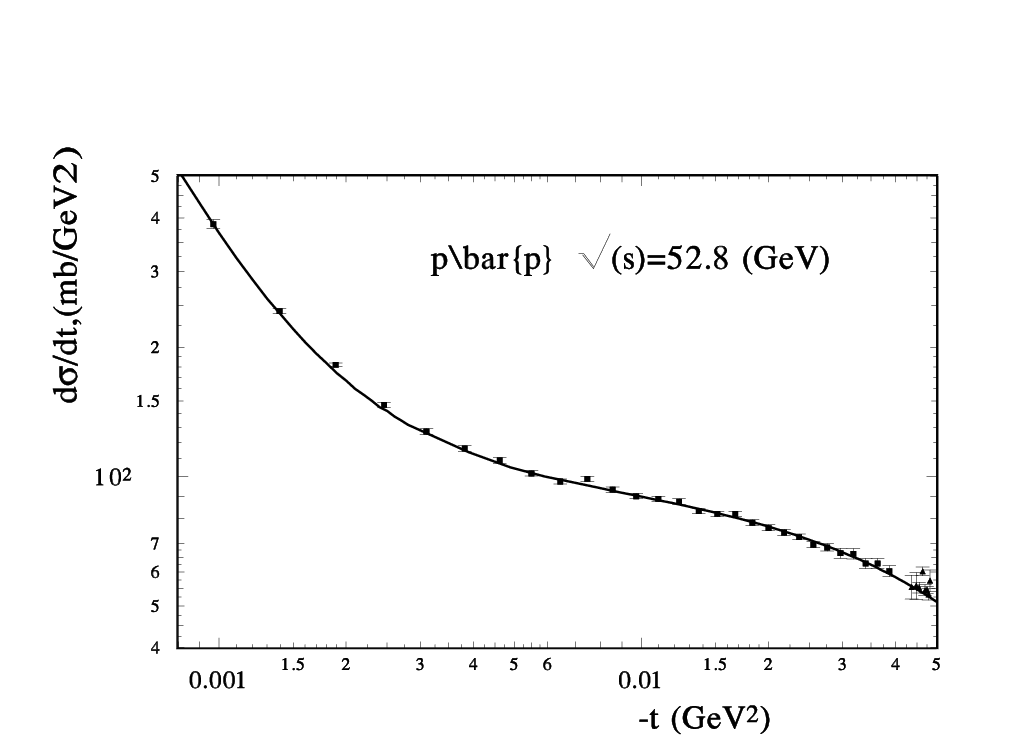}
\end{center}
\caption{ $d\sigma/dt \ {\rm at} \  \sqrt{s}=52.8 $~GeV and at small $t $ for $pp $
 (left) and for $p\bar{p}$ (right). }\label{Fig:MV}
\end{figure*}

 The final elastic  hadron scattering amplitude is obtained after unitarization of the  Born term.
    So first, we have to calculate the eikonal phase
   \begin{eqnarray}
 \chi(s,b) \   =  \frac{1}{2 \pi}
   \ \int \ d^2 q \ e^{i \vec{b} \cdot \vec{q} } \  F^{\rm Born}_{h}\left(s,q^2\right)\,,
 \label{tot02}
 \end{eqnarray}
  and then obtain the final hadron scattering amplitude
    \begin{eqnarray}
 F_{h}(s,t) = i s
    \ \int \ b \ J_{0}(b q)  \ \Gamma(s,b)   \ d b\,.
 \label{overlap}
 \end{eqnarray}
     \begin{eqnarray}
  \Gamma(s,b)  = 1- \exp[- \chi(s,b)] .
 \label{overlap}
\end{eqnarray}
   All these calculations are carried out by the FORTRAN program.

\section{Hadron form factors}
  As was mentioned above,  all the form factors are obtained from the GPDs of the nucleon \cite{ST-PRDGPD}.
The electromagnetic form factors can be represented as first  moments of GPDs
\ba
 F_{1}(t) = \int^{1}_{0}  dx  \ \sum_{u,d} {\cal{ H}}^{q} (x, t); \ \
 F_{2} (t) = \int^{1}_{0} dx \ \sum_{u,d} {\cal{E}}^{q} (x,  t),
\ea
following from the sum rules \cite{Ji97,R97}.

Let us modify the original Gaussian ansatz in order to incorporate
the observations of \cite{R98} and \cite{Burk04} and choose
 the $t$-dependence of  GPDs in the form
\ba
{\cal{H}}^{q} (x,t) \  = q(x) \   exp [  a_{+}  \
\frac{(1-x)^2}{x^{m} } \ t ].
\ea
  The value of the parameter $m=0.4$ is fixed by the low $t$  experimental data while
 the free parameters $a_{\pm}$ ($a_{+} $ - for ${\cal{H}}$
and $a_{-} $ - for ${\cal{E}}$) were chosen to reproduce the
experimental data in the whole $t$ region.  Indeed, large $t$ behavior
corresponds to $x \sim 1$ in (10), (11), where the dependence on $m$ is
weak.

The function $q(x)$ was chosen at the same scale $\mu^2=1$ as in \cite{R04},
which is based on the MRST2002 global fit \cite{MRST02}.
 In all our calculations we restrict ourselves, as in other quoted work,  to
 the contributions of $u$ and $d$ quarks.

 Hence, we have
\ba
 u(x) = 0.262 x^{-0.69}(1-x)^{3.50}(1+3.83x^{0.5}+37.65x),
\ea
\ba
 d(x) = 0.061 x^{-0.65}(1-x)^{4.03}(1+49.05x^{0.5}+8.65x).
\ea
 With this simple form
  we obtained a good description of the proton electromagnetic Sachs form factors.
  Using the isotopic invariance we obtained good descriptions of the neutron
  Sachs form factors without changing any  parameters \cite{ST-PRDGPD}.

 We shall use this model of GPDs to obtain the second momentum  form factor of the nucleon.
Taking instead of
the electromagnetic current $J^{\mu} $ the energy-momentum tensor $T_{\mu \nu} $ together
with a model of quark GPDs, one can obtain the gravitational form factor of fermions  \cite{ST-PRDGPD,ST-FF}
 For $\xi=0 $ one has
\ba
\int^{1}_{0} \ dx \ x \sum_{u,d}[{\cal{H}}(x,t) \pm {\cal{E}}(x,t)] = A_{h}(t) \pm B_{h}(t).
\ea
The integration of the second momentum of GPDs over $x$ gave  the momentum-transfer representation
  of the form factor (see Fig. 1). We approximate this  by the dipole form
\begin{eqnarray}
A(t)=L^{4}_{2}/(L^{2}_{2}-t)^2  .\label{overlap}
 \end{eqnarray}
with the parameter $L^{2}_{2}=2.0 \ $GeV$^2$.

\section{Fitting procedure}

The model has only three high energy fitting parameters
 and two low energy parameters, which reflect some small contribution
 coming from the different low energy terms.
 (see Table 1).
  We take all existing
      experimental data in the energy range $52.8 \leq \sqrt{s} \leq 1960 \ $GeV
      and the region of the momentum transfer $0.0008 \leq \ -t \ \leq 9.75 \ $GeV$^2$
      of the elastic differential cross sections of proton-proton and proton-antiproton
      data \cite{data-JR,data-Sp}.
       So we include the whole Coulomb-hadron interference region where the experimental errors are remarkably small. We do not include the data on total cross sections
        $\sigma_{tot}(s)$  and $\rho(s)$, as their values were obtained from the differential
        cross sections especially in the Coulomb-hadron interference region.
         Including such data decreases $\chi^2$. We also do not include the interpolated
         and extrapolated data of Amaldi \cite{Amaldi-166}.

   In the fitting procedure we calculate the minimum in  $\sum_{i=1}^{N} \chi_{i}^2$
   related with the statistical errors $\sigma_{i}^{2}$. The systematical errors
   are taken into account by the additional normalization coefficient $n_{k}$ for the $k$ series (the experiment) of the experimental data
\begin{eqnarray}
\chi^2= \sum_{i=1}^{N} \frac{n_{k} \ E_{i}(s,t) \ - \  T_{i}(s,t)}{\sigma_{i}^2(s,t)}  .\label{fit}
 \end{eqnarray}
 where  $T_{i}(s,t)$ are the theory predictions and $n_{k} E_{i}(s,t)$ are the  data points allowed to shift by the systematical error of the $k$-experiment (see, for example \cite{pdf-system1,pdf-system2}.

In the region of the small momentum transfer the systematic errors are of an order of $2 \% \div 5 \% $. For most part the additional normalization are in the region $0.95 \div 1.05$.
    At large momentum transfer the order of the systematical errors is
    $10 \% \div 20 \% $.
    In this case, the additional normalization is situated in the region $0.8 \div 1.2$.

\renewcommand{\bottomfraction}{0.7}

\label{sec:figures}
\begin{figure}[!b]
~\vspace{-1.cm}
\begin{center}
\includegraphics[width=0.4\textwidth] {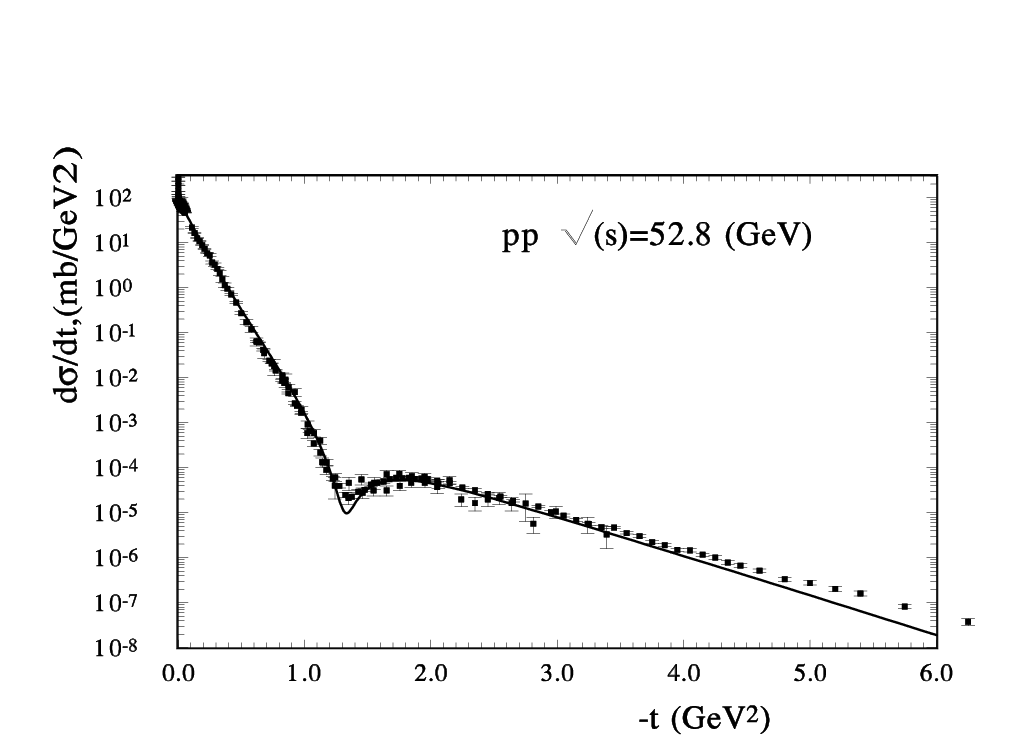}
\includegraphics[width=0.4\textwidth] {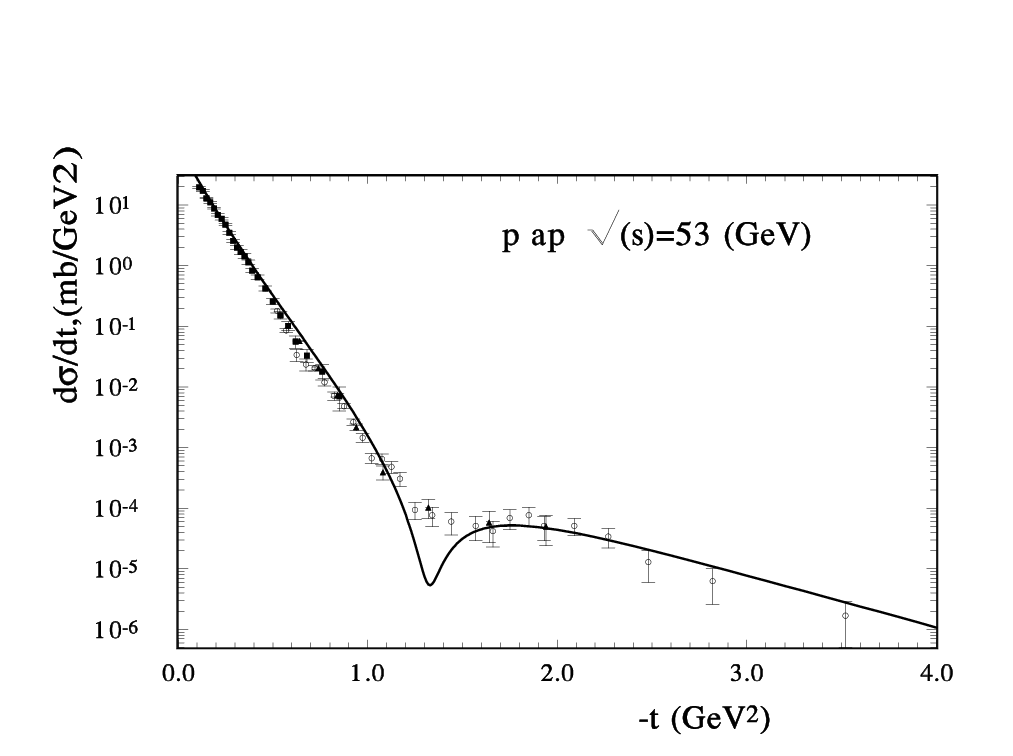}
\end{center}
\caption{$d\sigma/dt $ at $ \sqrt{s}=52.8 $~GeV  at large $|t| $ for $pp$ (top) and for $\bar{p}p$(bottom panel)
}\label{Fig:4}
\end{figure}

\label{sec:figures}
\begin{figure*}
\begin{center}
\includegraphics[width=0.4\textwidth] {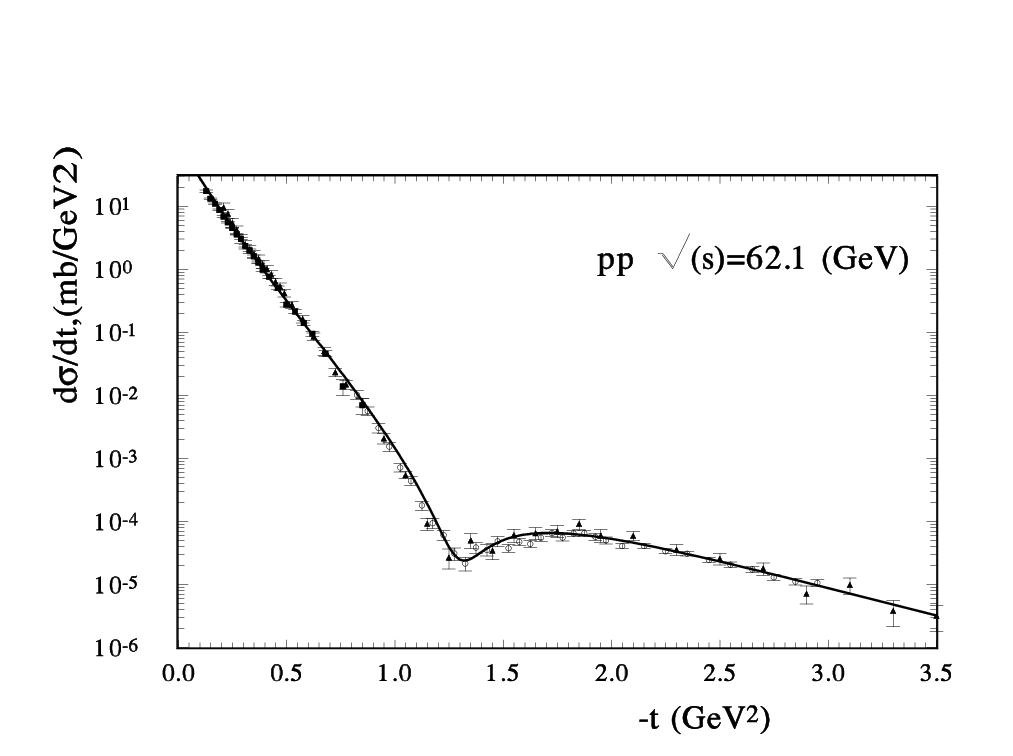}
\includegraphics[width=0.4\textwidth] {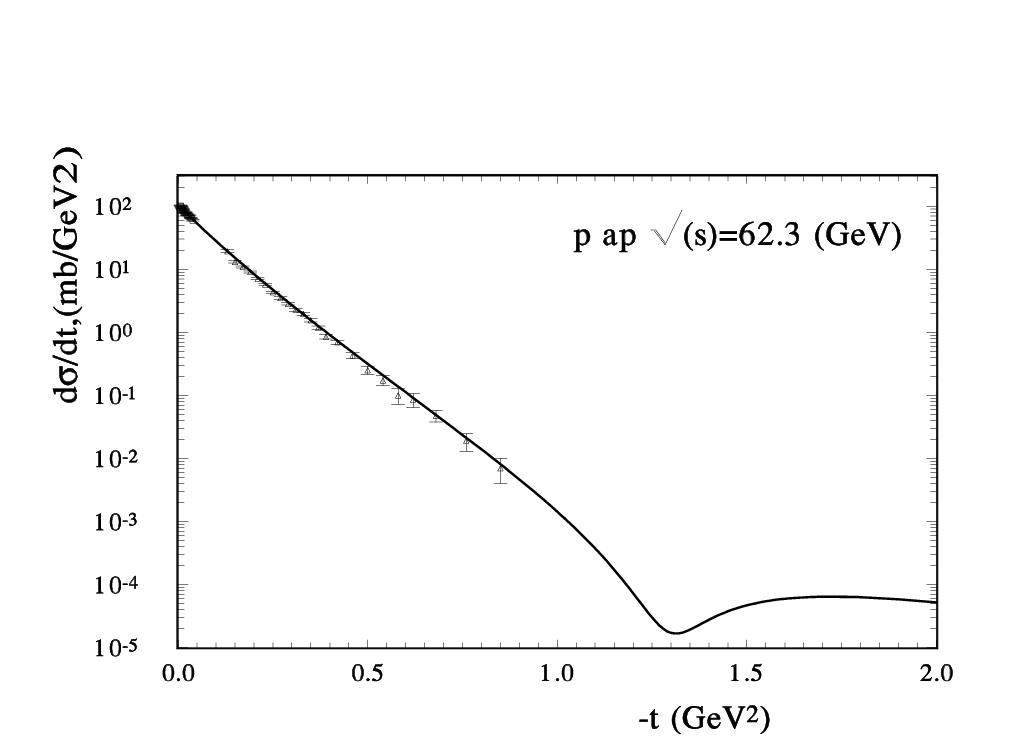}
\end{center}
\caption{ $d\sigma/dt $ at $ \sqrt{s}=62.1 $~GeV  at large $|t| $ for $pp$ (left) and for $\bar{p}p$(right panel)
}\label{Fig:4}
\end{figure*}

For the non-normalized experimental data of the $UA4/2$ Collaboration \cite{UA42} ,
      which have very small statistical errors,
       we take the normalization determined in \cite{Sel-UA42}. Our correction normalization
        is obtained from the fitting procedure in this case $n_{UA42}=0.95$.
\label{sec:figures}
\begin{figure}[!b]
\begin{center}
\includegraphics[width=0.4\textwidth] {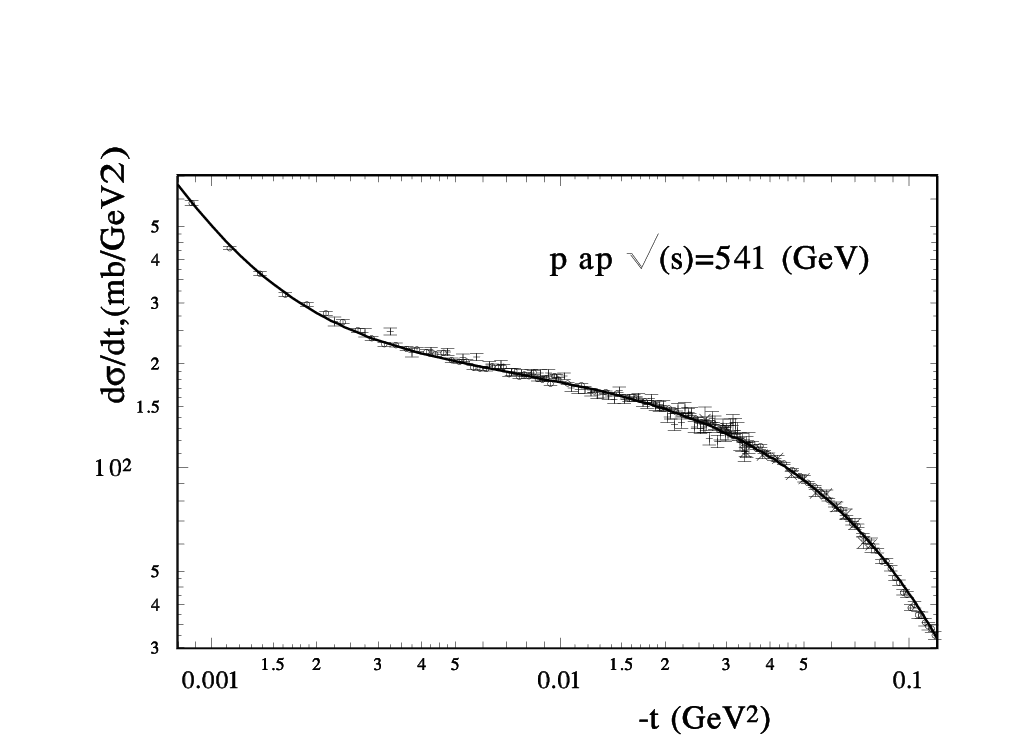}
\includegraphics[width=0.4\textwidth] {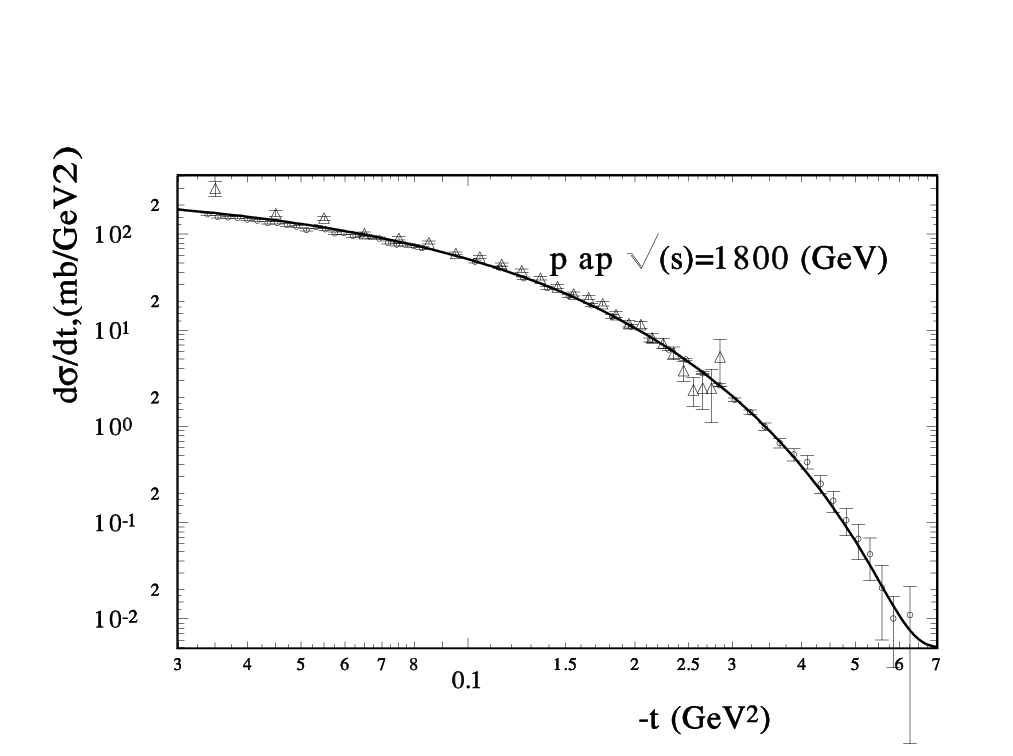}
\end{center}
\caption{ $d\sigma/dt $ for $\bar p p $ elastic scattering at small momentum transfer, at
 $\sqrt{s}=541 $~GeV (top) and
 $\sqrt{s}=1800 $~GeV (bottom)}\label{Fig:MV}
\end{figure}
\label{sec:figures}
\begin{figure}[!b]
\begin{center}
\includegraphics[width=0.4\textwidth] {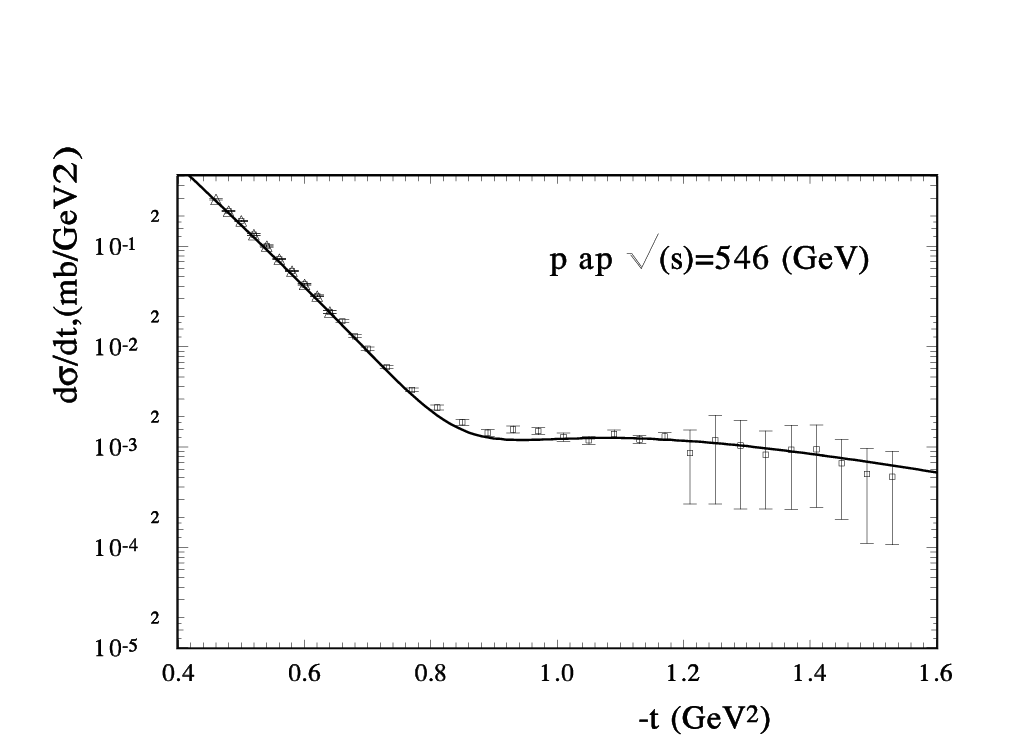}
\includegraphics[width=0.4\textwidth] {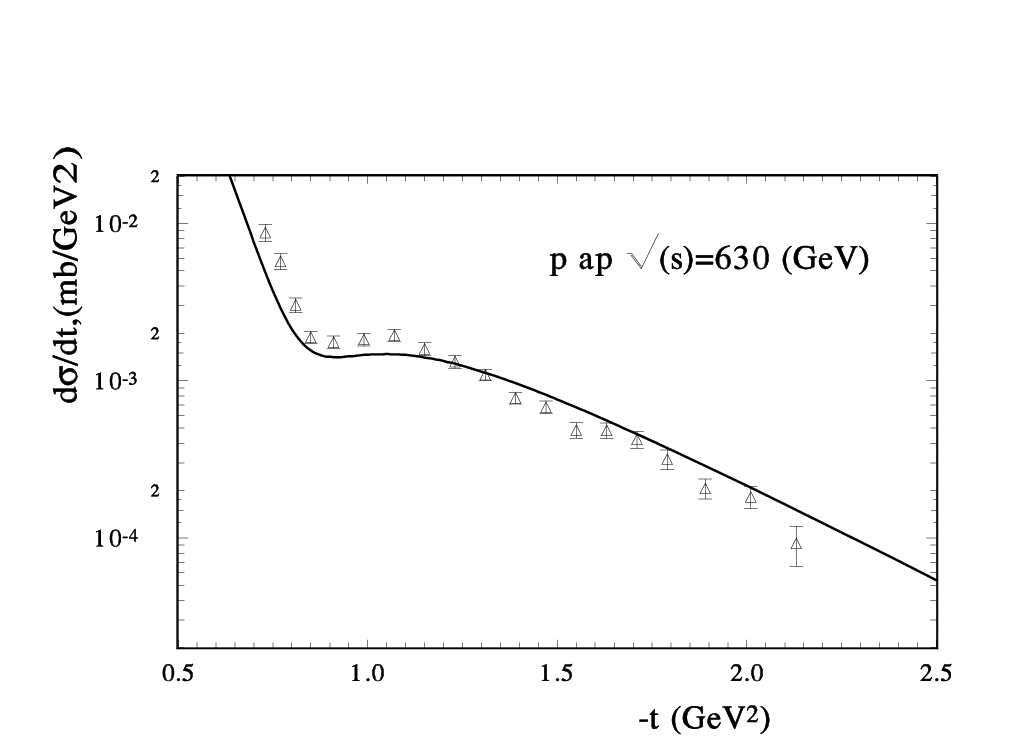}
\end{center}
\caption{ $d\sigma/dt $ for $\bar p p $ elastic scattering at large $t$, at
 $\sqrt{s}=546 $~GeV (top) and
 $\sqrt{s}=630 $~GeV (bottom panel)}\label{Fig:MV}
\end{figure}

\label{sec:figures}
\begin{figure}[!b]
\begin{center}
\includegraphics[width=0.4\textwidth] {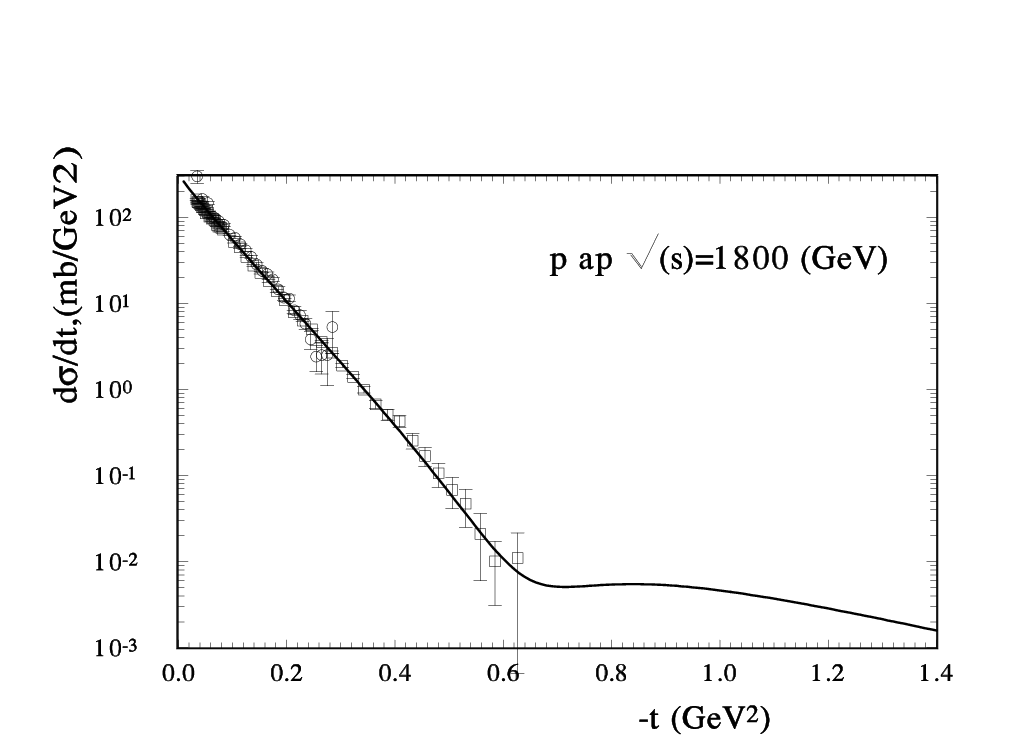}
\includegraphics[width=0.4\textwidth] {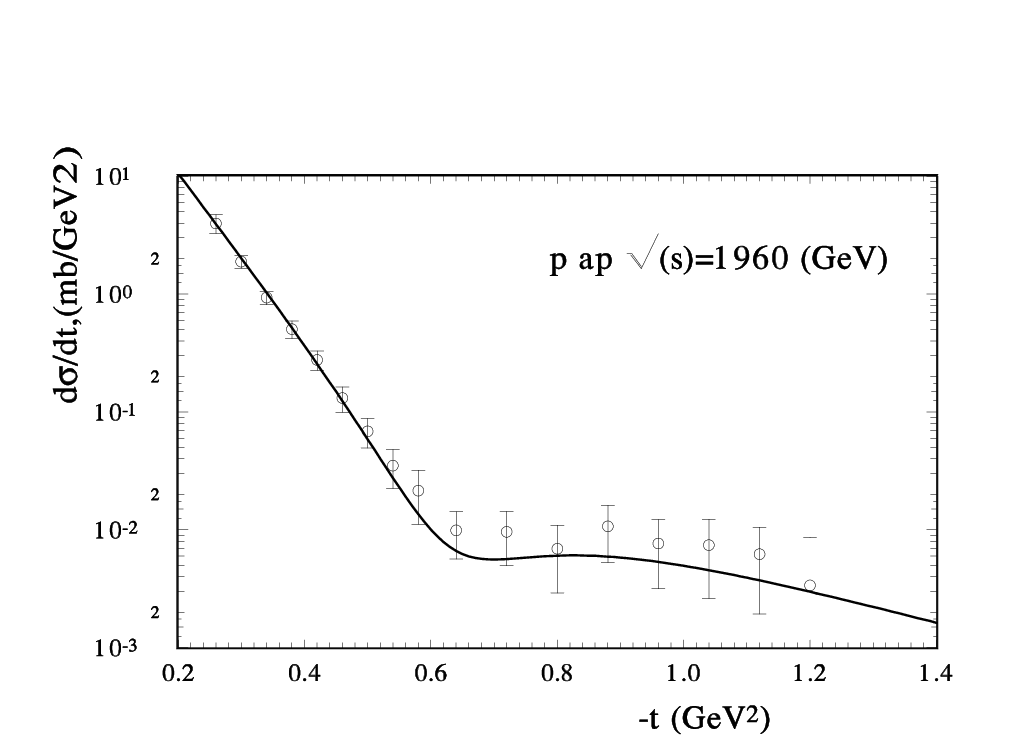}
\end{center}
\caption{ $d\sigma/dt $ for $\bar p p $ elastic scattering at large  $t$, at
 $\sqrt{s}=1800 $~GeV (top) 
 and
 $\sqrt{s}=1960 $~GeV 
   (bottom panel)}\label{Fig:MV}
\end{figure}

As a result, one obtains $\sum \chi^2_i /N \simeq 1.8 $ where $N=975 $ is
the number of experimental points.
 Of course, if one sums the systematic and
statistical errors, the $\sum \chi^2/N $ decreases, to $1.4 $.
Note that the parameters of the model are energy-independent.
The energy dependence of the scattering amplitude is determined
only by the single intercept and the logarithmic dependence on $s$ of the slope.

  Note that there are some separate points ( $n=17$)  at the different energies and
   momentum transfer  which give   $\sum_{n=1}^{17}  \chi_{n}^{2} =260$. However, we do not remove such points.

\section{Description of the differential cross sections}

 The differential cross sections for proton-proton elastic scattering at
 $\sqrt{s} = 52.8 $~GeV
are presented in Fig.~2(left panel) and 3(top panel). At this energy there are experimental data at small
(beginning at $-t=0.001 $~GeV $^2 $) and large (up to $-t=10 $~GeV $^2 $) momentum transfers.
The model reproduces both regions and provides a qualitative description of the dip region
at $-t \approx 1.4 $~GeV$^2 $, for $\sqrt{s}=53 $~GeV$^2 $ and for $\sqrt{s}=62.1 $~GeV$^2 $ (Fig.3(top) and Fig.4(lett panels)).

Now let us examine the proton-antiproton differential cross sections (see Fig. 2(right panel)).
In this case at small momentum transfer
the Coulomb-hadron interference term plays an important role and has the opposite sign.
The model describes the experimental data well.
Slightly worse than $pp$ is the description of  $p\bar{p}$ of
  differential cross sections at $\sqrt{s}=53 \ $GeV, especially in the diffraction minimum (Fig. 3(bottom panel)).

\label{sec:figures}
\begin{figure}
\begin{center}
\includegraphics[width=0.4\textwidth] {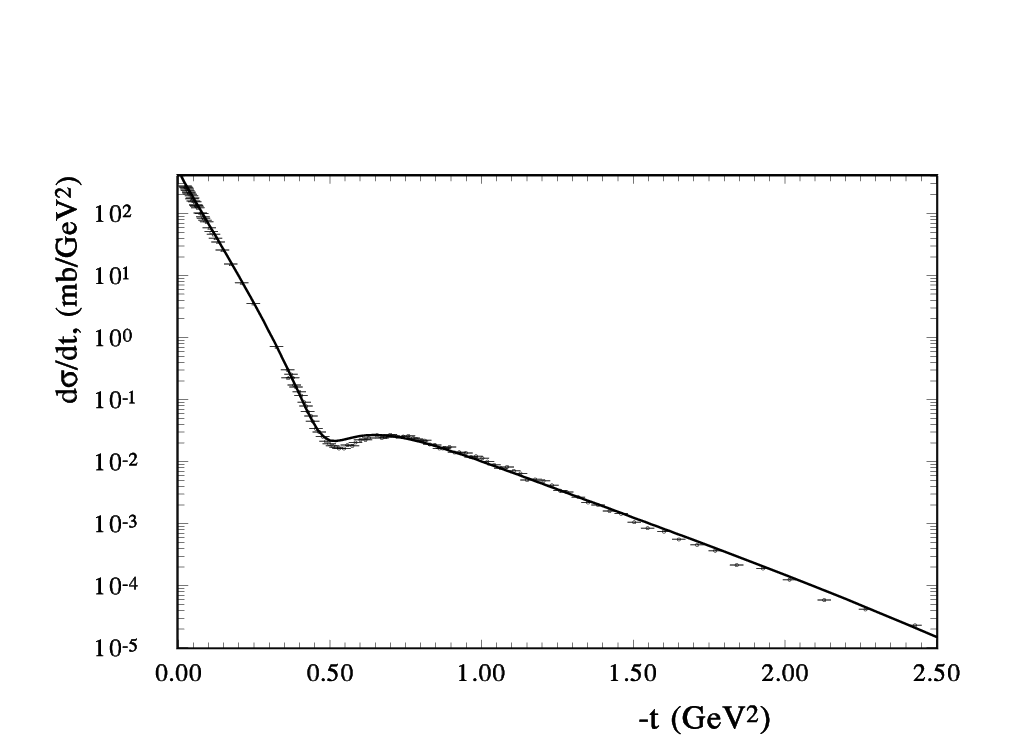}
\end{center}
\caption{The comparison of the model calculations with the parameters, based on the fit of the
  experimental data at $52.8  \leq \sqrt{s} \leq  1960 \ $GeV,
 with the preliminary data of the TOTEM Collaboration at
 $\sqrt{s}=7 $~TeV.
 }\label{Fig:MV}
\end{figure}
\label{sec:figures}
\begin{figure}[!b]
\begin{center}
\includegraphics[width=0.4\textwidth] {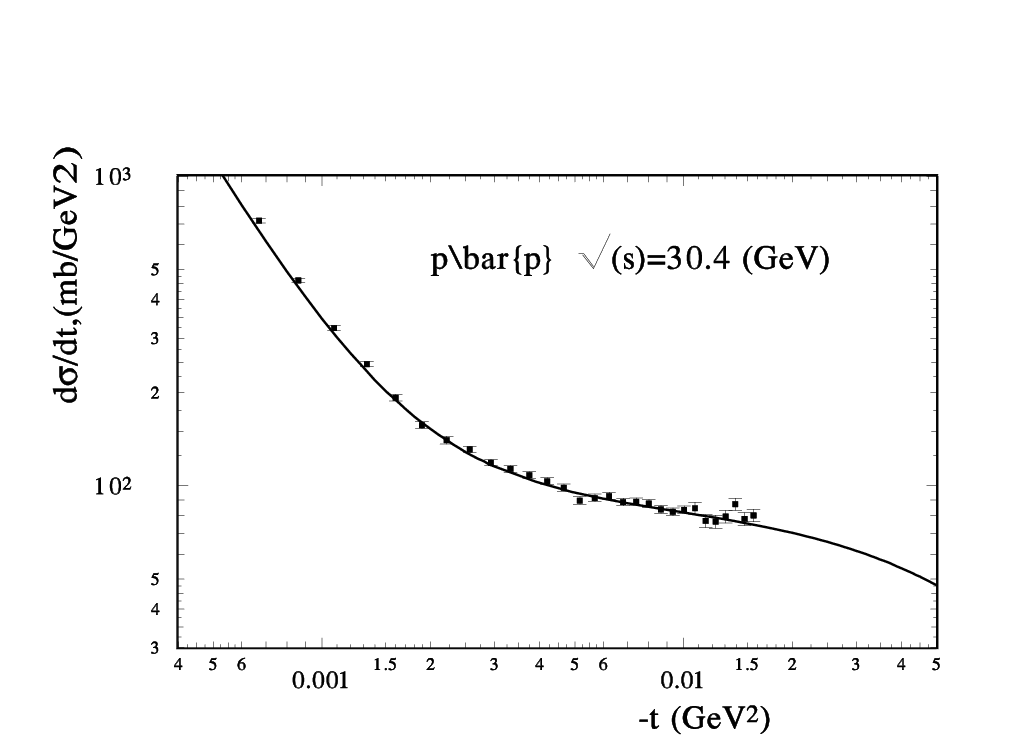}
\includegraphics[width=0.4\textwidth] {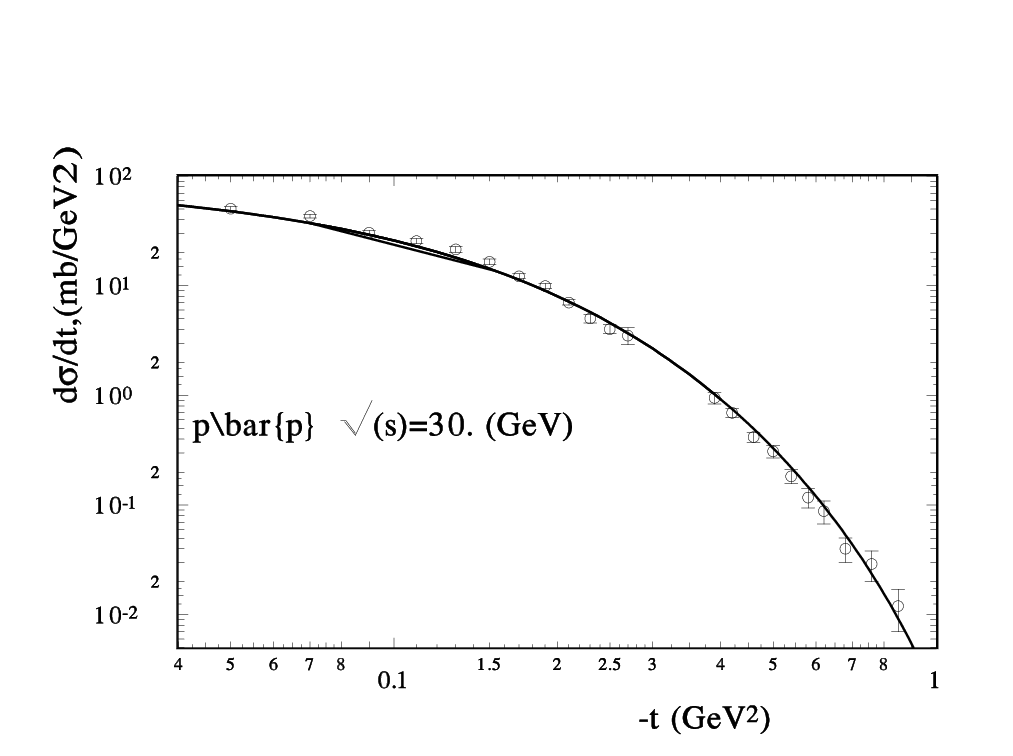}
\end{center}
\caption{The model predictions of $d\sigma/dt $ for $\bar p p $  elastic scattering at
 at  $\sqrt{s}=30.6 $~GeV.}\label{Fig:ss30}
\end{figure}

  Maybe, this shows an additional  odderon contribution. Note that at $\sqrt{s} = 62.2 \ $ GeV
  for $p\bar{p}$ scattering the description of the differential cross sections is essentially better (Fig.~4).

In Fig.~5, the description of the proton-antiproton scattering at $\sqrt{s}=541 \ $~GeV and at
 $\sqrt{s}=1800 \ $~GeV is shown.
 In this cases, the Coulomb-hadron interference term is large, especially at $\sqrt{s}=541 \ $~GeV as $t$ is very small.
The good description of the experimental data shows that the
  energy dependence of the real part of scattering amplitude obtained in the model  corresponds to the real physical situation.

Figures 6 and 7 show the description of the experimental data at larger momentum transfers for
 $\sqrt{s}=546 $~GeV$^2 $ and $\sqrt{s}=630 $~GeV$^2 $ and for Tevatron energies
  $\sqrt{s} = 1800$ GeV and $\sqrt{s} = 1960 \ $GeV.  It is clear that the model leads to a
good description of these data in the region of the diffraction minimum without taking into account
 the odderon contribution. Hence, it is shown that  very likely the intercept of the odderon is near $1$.

 On basis of this fit of the experimental data
 at $52.8  \leq \sqrt{s} \leq  1960 \ $GeV and $0.0008 \leq |t| \leq  9.75 \ $GeV$^2$ we obtained the fitting parameters (see Table 1). Taking into account these values of the parameters
 we extend the scope of the model  and  calculate the
 differential cross sections at $7 \ $TeV for $pp$ elastic scattering. In Fig. 8, the comparison of
 the model calculations with the parameters, based at the fit of the existing experimental data at $52.8  \leq \sqrt{s} \leq  1960 \ $GeV,
 with the preliminary data of the TOTEM Collaboration are shown. Except the size of the diffraction minimum the coincidences are remarkable.   Of course, if we include in the model some different correction terms,
  like odderon, the value of the fitting parameters of the model will be slightly change. However, we think that
  the basic properties of the model will not change in future.

\label{sec:figures}
\begin{figure*}
\begin{center}
\includegraphics[width=0.4\textwidth] {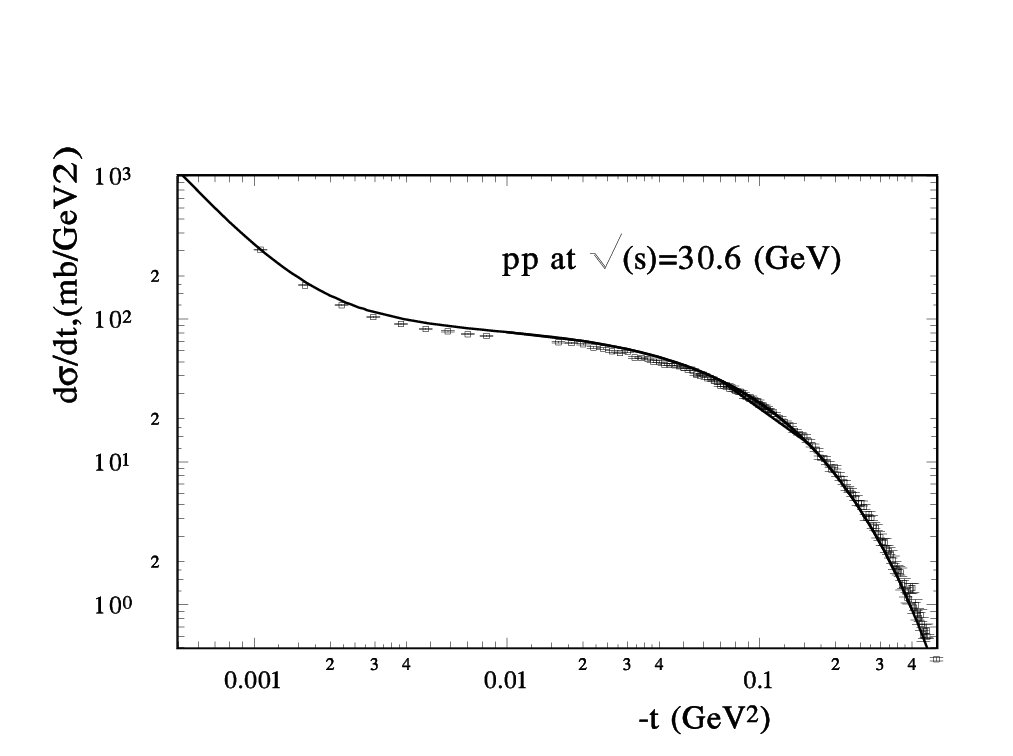}
\includegraphics[width=0.4\textwidth] {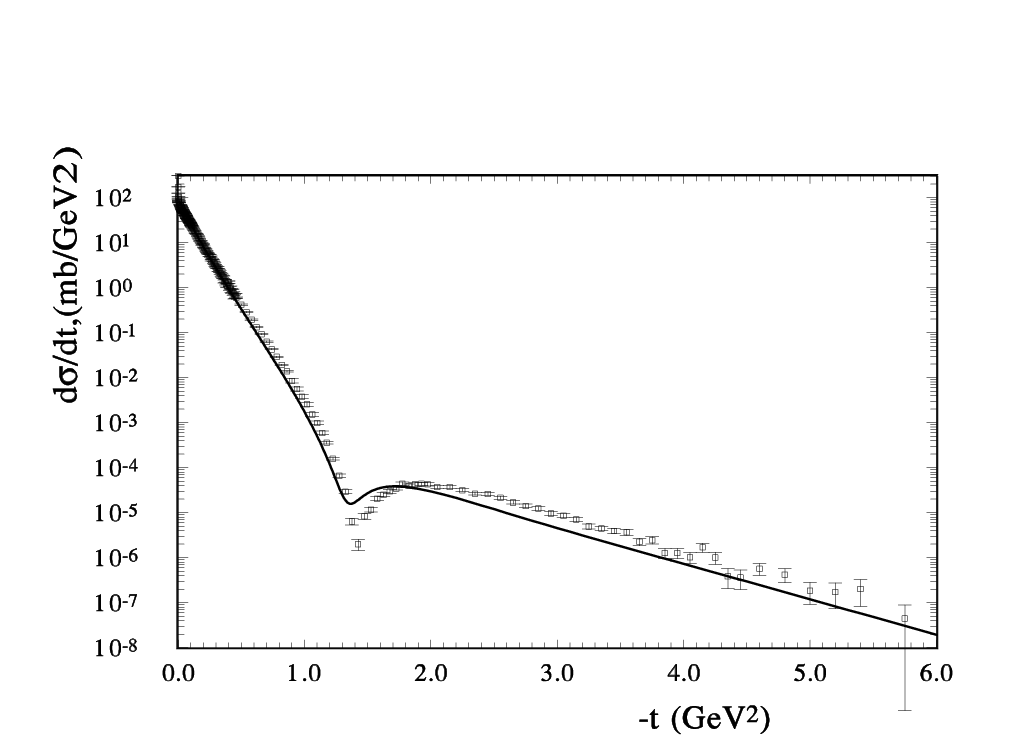}
\end{center}
\caption{ The model predictions of $d\sigma/dt $ for $ p p $ elastic scattering at
 at  $\sqrt{s}=30.6 $~GeV.}\label{Fig:ss30}
\end{figure*}

\label{sec:figures}
\begin{figure}[!b]
\begin{center}
\includegraphics[width=0.4\textwidth] {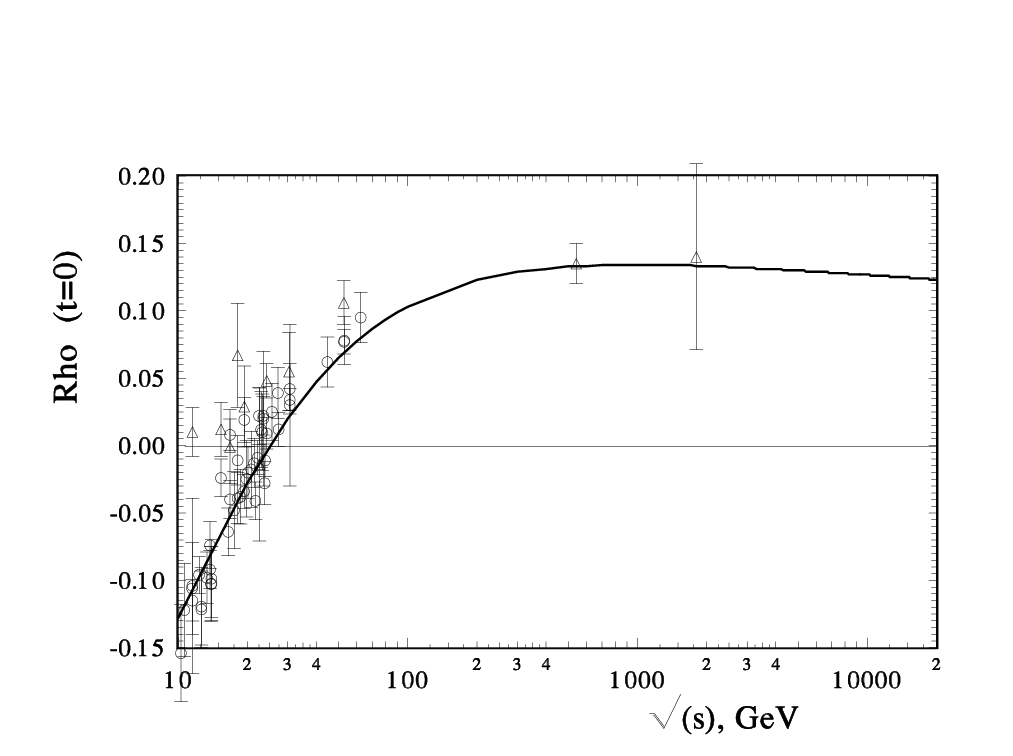}
\includegraphics[width=0.4\textwidth] {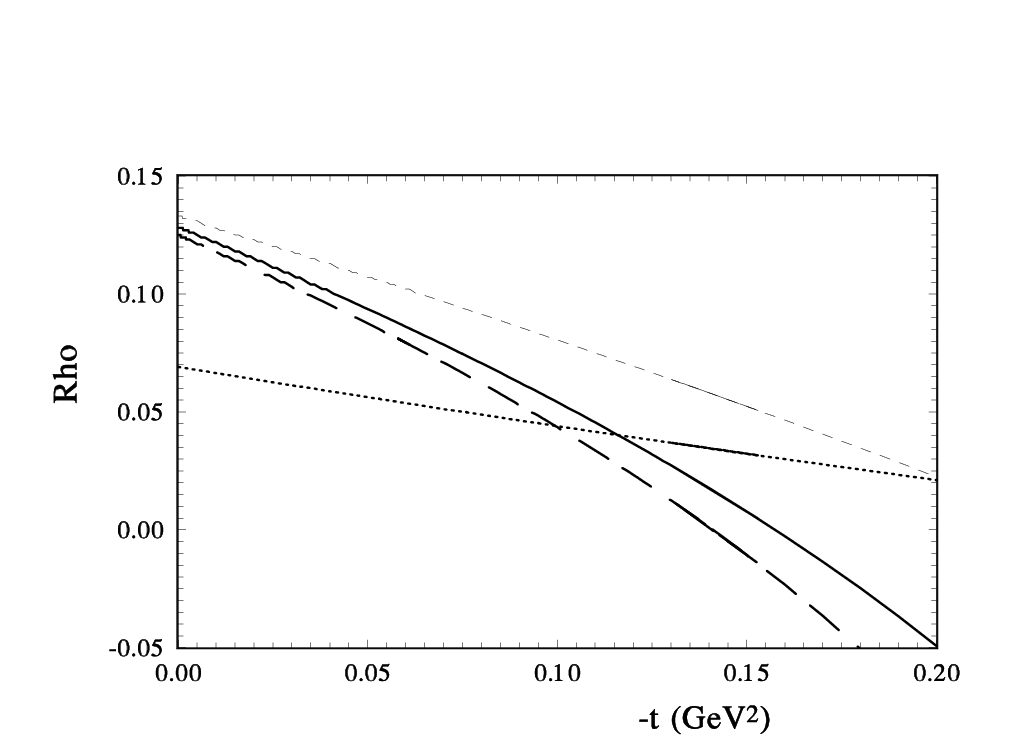}
\end{center}
\caption{  $\rho(s,t)$ for $p p $ elastic scattering at $t=0$ (top);
 (circle and triangles are the experimental data of $pp$
  and $\bar{p}p$ elastic scattering, respectively.
 and  at small momentum transfer (bottom) at
 $\sqrt{s}=52.8 \ $~GeV (tiny-dash line), $\sqrt{s}=541 $~GeV (dash line),
 $\sqrt{s}=7 \ $~TeV (hard line), and $\sqrt{s}=14 \ $~TeV (long-dash line)   }   \label{Fig:rho}
\end{figure}
\label{sec:figures}
\begin{figure}
\includegraphics[width=0.4\textwidth] {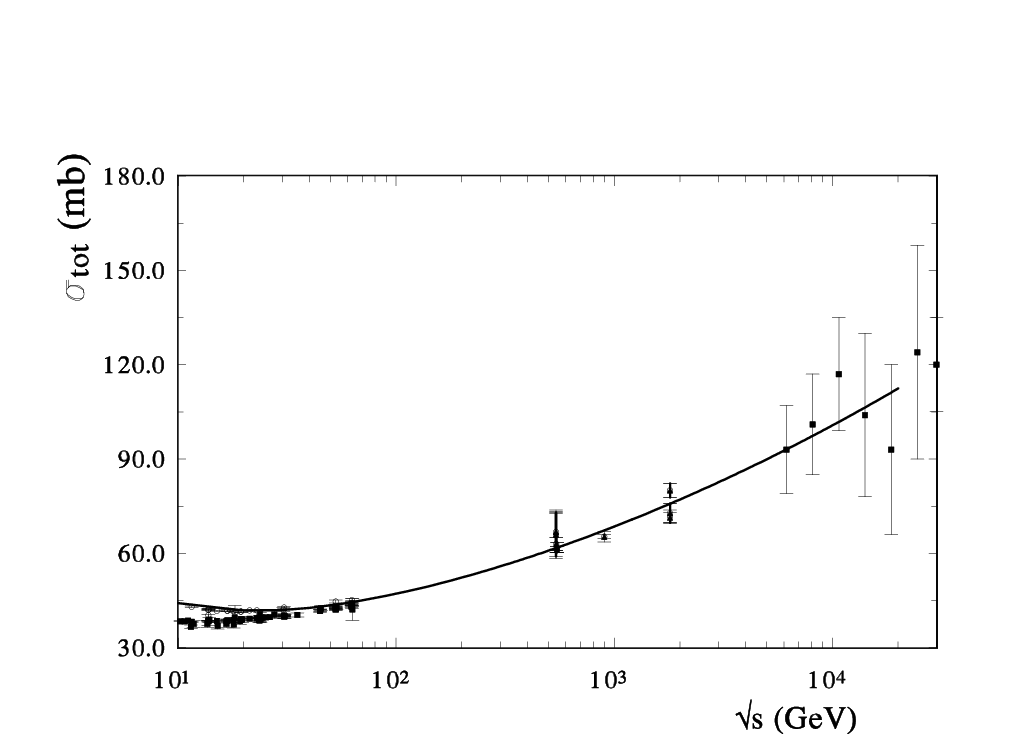}
\caption{$\sigma_{tot}(s)$ are calculated in the model (hard line).}
 (circle and triangles are the experimental data of $pp$
  and $\bar{p}p$ elastic scattering, respectively.
\end{figure}

Now let us see how
 we can  extend the scope of the model to a low energy .
The calculation of the model of $d\sigma/dt $ for $\bar p p $  and $pp$  elastic scattering
 at  $\sqrt{s}=30.6 $~GeV are compared with the experimental data on Fig.9 and Fig.10.
 There is a very good description at small momentum transfer for  both the reactions
  - $pp$ and $\bar{p}p$. At large $t$ the model reproduces the differential cross sections
   only qualitatively. The position of the diffraction minimum corresponds to the experimental data.
   However, the form of the diffraction minimum obviously requires some additional small
   correction terms, possibly by the odderon contributions. This situation repeated
   the problem of describing the form of the diffraction minimum at $\sqrt{s} = 53 \ $GeV for $p\bar{p}$
    elastic scattering.

\section{Energy dependence of   $\rho(s,t)$ and  $\sigma_{tot}(s)$}

     The ratio  of the real part to the imaginary part of the elastic scattering hadronic amplitude
     $$\rho(s,t) = Re F_{h}(s,t)/Im F_{h}(s,t)$$ is
  very important as it reflects the $t$-dependence of the both parts of the scattering amplitude,
    which are connected one to the other through
   the integral dispersion relations.
The validity of this relation can be checked at LHC energies. The deviation can point
      out  the  existence of a fundamental length at TeV energies \cite{Khuri1,Khuri2}.
 Usually, the value of $\rho(s,t=0$
is assumed to be small and to vary little with $t$: $\rho(s,t)\approx 0.14$. 
   The differential cross sections at small momentum transfer $  |t| \leq 0.05 \ $GeV$^2$,
   the so-called Coulomb-hadron interference region, are determined by the interference of the Coulomb
   amplitude with the real part of the hadron amplitude. Hence, the $s$ and $t$ dependence
   of the  real part of the hadron amplitude, which is reflected in  $\rho(s,t)$, will
   determine the form of the differential cross sections.

   In the model, the real part
   of $F_{h}(s,t)$ is determined only by the complex cross-symmetric form of energy $\hat{s}=s \ exp[-i \pi/2]$.
    No other artificial function or some parameters
   impact  the form or $s$ and $t$ dependence of the real part.
     Despite such simplicity, the model sufficiently well describes the experimental
     data in the Coulomb-hadron region of momentum transfer and in a wide energy region
     (see Figs. 2,5,9). The calculated $\rho(s, t=0)$ in the  model is shown in Fig. 11(top panel).
     For the most part it coincides with  $\rho$ obtained by the COMPETE Collaboration
     \cite{COMPETE1,COMPETE2}. We can see that the $\rho(s,t)$ reaches its maximum $0.135$ at approximately
     $1 \ $TeV and then slowly decreased up to $0.127$ at $\sqrt{s} = 20 \ $TeV.
    Note that the model takes into account the only cross-symmetric part of the scattering amplitude.
    This  reflects that at low energy it, in most part, coincides with the
    experimental data of the proton-proton scattering.
       The $t$ dependence of $\rho$, shown in Fig. 11(bottom panel),  is interesting.
      We can see that after $\sqrt{s} = 1 \ $TeV, where $\rho$ reaches its maximum,
       its $t$-dependence is changed. If the maximum of $\rho$ is small decreasing, his slope
       will be larger and larger. It is interesting to compare this figure with
        a figure like this in our phenomenological analysis of $\rho$ at high energies \cite{CS-PRL09}.
        The size of $\rho$  is essentially larger, but the dynamic change of $\rho$ with $s$ and
       $t$ is practically the same. It is related with the saturation regime which impacts
       on the $s$ and $t$ dependence of $\rho(s,t)$ at high energies.

       As  the data on the total cross section $\sigma_{tot}(s)$ and $\rho(s,t)$ were not included in our
   fitting procedure, let us now calculate these values in the framework of the model.
   The energy dependence of the $\sigma_{tot}(s)$ is shown in Fig. 12. Our model calculations
   coincide with the existing experimental values sufficiently well. The model calculations gives
   $\sigma_{tot} = 43.46, \ 61.7, \ 75.76 \ $mb for the energies - $\sqrt{s} = 50, \ 540, 1800 \ $GeV.

\section{Conclusion}

    We present the new model of the hadron-hadron interaction at high energies.
  The model is very simple from the viewpoint of the number of parameters and functions.
  There are no any artificial functions or some cuts which bound the separate
  parts of the amplitude by some region of momentum transfer.
  The model describes all high energy data sufficiently well, except the form of
  the diffraction minimum of the proton-antiproton scattering at $\sqrt{s} =53 \ $GeV.
  As we know, it is the only  model which describes all available high energy data
  in the Coulomb-hadron region and large momentum transfer.
   The energy dependence of the differential cross sections is determined  by only one intercept
  $\Delta = 1 + \epsilon $ with $\epsilon =0.11$ and the $\alpha^{\prime} =0.24 \ ln[\hat{s}]$.
  One of the most remarkable properties is that the real part of the
  hadron scattering amplitude is determined only by complex energy $s$ that satisfies the crossing-symmetries.
  The differential cross sections at small momentum transfer are determined, in most part,  by the Coulomb-hadron
  interference term.
    This is essentially caused by the $s$ and $t$ dependence
 of the real part of the hadronic amplitude. A good description of the differential cross section (see figs 2, 5, 9) confirms
  the determination of the real part in the model.

     However, the most important advantage of the model is that it is built on some  physical
     basis - two form factors which are calculated from GPDs.
The behavior of the differential cross
section at small  $t $ is determined, in most part,  by the electromagnetic form factors; and  at large  $t $,
 by the matter distribution (calculated in the model from the second momentum of the GPDs),
 as was supposed by H. Miettinen a long time ago~\cite{Miettinen} and then used in the the work
  of S. Sanielevici and P. Valin \cite{Valin}.
   Of course, we understand that this advantage of the model, on the other hand, has
   some disadvantage. Now we have no rigorous proof of such a physical picture.
   However, the best work of the model maintains the hope that such a representation has a right    to exist.


  The model predicts $\sigma_{tot}=95$ mb at $\sqrt{s}=7$ TeV. The preliminary data of the TOTEM show
     $\sigma_{tot}=98.3 \pm 0.2^{stat.} \pm 2.8^{syst.} $ mb \cite{TOTEM-1395}.
Of course,  some correction terms (corresponding to the odderon or spin-flip amplitudes)
     have to be included in our model.
    Now the model does not show a contribution of the hard pomeron in the examined energy
    region. The presence of the hard pomeron has to give an additional contribution in the
    real part of the scattering amplitude and change the size and form of the Coulomb-hadron
    interference term \cite{CS-PRL09}. Hence, we hope that all these additional terms  will be determined after fitting with the new data of the proton-proton scattering at LHC energies.

\vspace{0.5 cm}

{\bf Acknowledgements}:  {\small The author would like to thank  J.-R. Cudell for helpful discussions,
 gratefully acknowledges  the financial support
  from BELSPO and would like to  thank the  University of Li\`{e}ge
  where part of this work was done.
   }

\begin{footnotesize}

\end{footnotesize}
\end{document}